\documentclass[
reprint,
superscriptaddress,
amsmath,amssymb,
aps,
]{revtex4-2}

\usepackage{graphicx}
\usepackage{dcolumn}
\usepackage{bm}

\begin{document}

\preprint{APS/123-QED}

\title{Multimode interface between optical free-space- and waveguide modes}

\author{Teresia Stranden}
\affiliation  {Tampere University, Photonics Laboratory, Physics Unit, 33720 Tampere, Finland}
\author{Oussama Korichi}
\email{oussama.korichi@tuni.fi}
\affiliation  {Tampere University, Photonics Laboratory, Physics Unit, 33720 Tampere, Finland}
\author{Matias Eriksson}
\affiliation  {Tampere University, Photonics Laboratory, Physics Unit, 33720 Tampere, Finland}
\author{Matteo Cherchi}
\affiliation  {current affiliation: Xanadu Quantum Technologies Inc.
777 Bay St., Toronto, ON M5B 2H7, Canada}
\affiliation  {VTT Technical Research Centre of Finland Ltd, Tietotie 3, 02150 Espoo, Finland}
\author{George Thomas}
\affiliation {VTT Technical Research Centre of Finland Ltd, Tietotie 3, 02150 Espoo, Finland}
\author{Robert Fickler}%
\affiliation  {Tampere University, Photonics Laboratory, Physics Unit, 33720 Tampere, Finland}

\begin{abstract}
Free-space and on-chip photonic systems are key components in optical communication networks. While free-space beams allow for the flexible generation and manipulation of spatial modes, integrated waveguides provide compact and stable platforms for on-chip signal processing. Bridging these two domains is essential for scalable multi-mode communication networks. Here, we present an efficient, broadband interface capable of converting multiple higher-order free-space Laguerre–Gauss (LG) modes into corresponding waveguide modes using the multi-plane light conversion (MPLC) scheme. We experimentally demonstrate low-crosstalk mode conversion between various set of three LG modes, and the first three TE modes of a multimode silicon waveguide across the telecom C-band. The system operates passively without active switching and can be adapted to different spatial mode sets. This platform provides a pathway to increased data capacities and may enable more compact and efficient multi-mode optical communication and on-chip processing schemes.
\end{abstract}

\maketitle

\section{\label{sec:level1}Introduction}

With an ever increasing need for higher data rates, photonics has become an important player in high-speed data transfers and on-chip information processing. 
Here, a key focus has been increasing the ability to multiplex multiple signals using various frequency \cite{keiser1999review, liu2009wavelength, jorgensen2022petabit} and polarization \cite{ivanovich2018polarization} bands with mostly only using a single spatial channel, e.g., a single-mode fiber or waveguide. As first signs of reaching the fundamental limits of wavelength multiplexing have been achieved \cite{hamaoka2018150}, the focus has been widened to also explore additional routes to avoid a so-called capacity-crunch, namely the focus on additional multiplexing using spatial modes. 

Over long distances higher-order spatial modes in fiber and free space have been explored to enlarge the number of multiplexed channels achieving data rates up to Peta-bit-per-second in either domain \cite{rademacher2021peta, zhou2025demonstration}. 
In free space, Laguerre-Gauss (LG) modes have widely been used as the preferred set of traverse spatial modes \cite{al2021structured}, as their cylindrical symmetry nicely matches most optical systems, and thus they allow the highest information density for a given aperture if the full transverse domain (azimuthal and radial) is taken into account \cite{zhao2015capacity}.
Standard higher-order fiber modes show a similar transverse spatial profile, however, fiber stress and bends commonly result in complex mode-coupling dynamics during the propagation of optical signals, which can either be mitigated through custom-tailored fibers  \cite{brunet2015design} or the use of a topologically protected subset of modes \cite{ma2023scaling}.
Specialized multi-mode fibers have further been explored to realize amplifiers working simultaneously for multiple modes along with studies exploring nonlinear optical effects in such advanced fiber systems \cite{cristiani2022roadmap}. 
Over short distances of signal transmissions and for local optical signal processing tasks, multi-mode waveguide chips have been developed to also provide an enhanced solution for photonic integrated circuits (PIC). 
Here, significant progress has been made in solving the challenges of effective mode manipulations and the beneficial use of the offered increased flexibility in the design of such devices, incl. on-chip multiplexers, active mode manipulations, and mode couplers to name a few \cite{cai2012integrated,li2019multimode,zhang2025fully}.
With fiber- and free space spatial modes being the backbone for long-distance transmission of high-bandwidth optical communication and multimode silicon photonic chips serving as local high-capacity information processing units, a natural requirement in future multimode networks appears: 
An efficient interface between sets of free space and waveguide modes that simultaneously links all data channels, ideally over a broad frequency range not to prevent additional wavelength division multiplexing.\\
\begin{figure}[ht]
\centering
\includegraphics[scale=0.48]{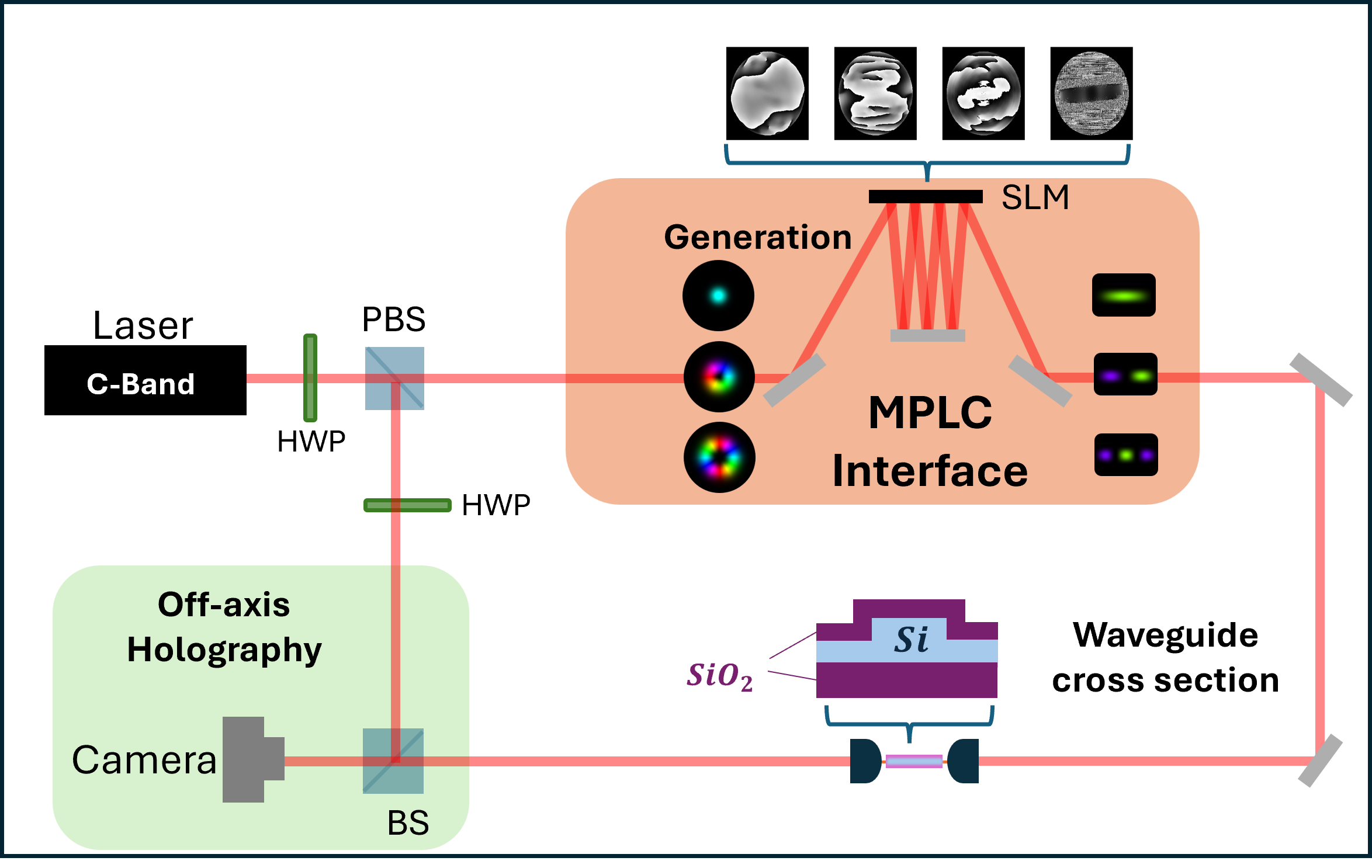}
\caption{
Sketch of the experimental setup. ($LG_{00}$, $LG_{10}$, $LG_{20}$) modes are generated using the first hologram. Then, the multi-plane light conversion (MPLC) system is implemented with a spatial light modulator (SLM). The MPLC converts the input modes to the first waveguide modes, which are coupled into a multimode Si rib waveguide using an aspheric lens. After propagation through the waveguide, the output field is imaged onto a camera. Off-axis digital holography is performed by interfering the signal beam with a reference beam, which is separated before the MPLC and adjusted in intensity and polarization using waveplates. A similar interferogram is recorded before the waveguide to characterize the MPLC output.}
\label{figure1}
\end{figure}
In this letter, we present a multi-mode interface that is able to convert multiple free-space optical modes into waveguide modes over the full C-band without any need for active switching between the channels. 
Our interface is based on a multi-plane light conversion (MPLC) scheme, which is realized by four consecutive phase modulation masks with a short propagation distance between. 
We show that a set of three higher-order Laguerre-Gauss modes can be converted to the first three higher-order waveguide modes of a silicon multi-mode waveguide without inducing a significant crosstalk. 
We further show that the converted light can be efficiently coupled and transmitted to the multi-mode waveguide, again without significant increase of crosstalk and for the entire telecom C-band. 
Finally, we demonstrate that the interface can also be adjusted to different sets of LG modes including their higher radial orders without a change in performance. Given the similarity of higher fiber- and free-space modes, our device can be seen as an essential building block for future multi-mode optical communication and data processing. 

\section{Background}
At the heart of our work lies an MPLC system, which is a scheme that was recently developed for spatial mode multiplexing \cite{labroille2014efficient,fontaine2019laguerre,rademacher2021peta} but also found its applications in single-to-multi fiber coupling \cite{korichi2023high}, image processing \cite{rouviere2024ultra}, sensing \cite{eriksson2023sensing}, turbulence characterization in free space transmissions \cite{billaud2022free}, quantum gate operations \cite{brandt2020high}, and quantum state characterization \cite{lib2022processing} to name a few of many recent applications. 
In our implementation, the MPLC is used to realized a interface between a set of free space transverse spatial modes and a multi-mode waveguide, i.e., it is designed to convert a given set of LG modes to the modes of a rib waveguide. 
In this proof-of-principle demonstration we restrict our MPLC scheme to be realized by only four phase modulations resulting in the limitation of being able to convert a set of three modes simultaneously. 
The required phase modulations are obtained using an optimization algorithm called wavefront matching \cite{hashimoto2005optical} and is implemented using a liquid-crystal phase-only spatial light modulator. 
We note that more evolved modulation schemes with more modulation planes have been demonstrated, meaning that the interface’s performance could be improved in terms of efficiency, cross-talk, and also number of converted modes and, Thus, can be scaled up to interface a larger subset of waveguide modes supported by the chip than those addressed in this work.

\section{Experimental setup}
The experimental setup is sketched in Fig. \ref{figure1}. 
A laser beam from a fiber-coupled continuous-wave tunable laser (ID Photonics CoBrite DX1 S, $\lambda~\in~[1528~\text{nm}, 1565~\text{nm}]$) is first collimated and magnified with a 4-f system to a beam waist radius of $3.0$~mm, which then enters the MPLC system.

The four required phase profiles for the MPLC are displayed on a single spatial light modulator (Holoeye Pluto-2) on different spatial locations. The beam is reflected between the SLM and a mirror impinging on the SLM four times, applying each of the phase modulations upon consecutive reflections from the SLM.
The first hologram is additionally used to generate the input set of spatial modes of light with arbitrary amplitude and phase via holographic shaping techniques \cite{Bolduc:2013aa}.
The output of the MPLC is demagnified by a factor of $\approx130$ 
with two cascaded 4f systems
, imaging the beam onto the input facet of a multimode waveguide.

The waveguides are fabricated on Silicon-on-Insulator (SOI) wafers with 150~$mm$ diameter. Here the device layer thickness is 3 $\mu m$, and the platform is generally referred to as a thick-SOI platform \cite{aalto2019open,cherchi2023supporting}. The fabrication process involves a UV stepper process, silicon waveguide etching based on the Bosch process \cite{gao2014smooth}, and cladding deposition. With this process, rib waveguides with a width of 26.9~$\mu$m are created by partially etching the silicon layer by 1.2~$\mu$m.

The field propagates through the $5$~mm -long waveguide, after which the light field at the output facet is imaged with magnification factor $\approx 195$ using two cascaded 4f systems %
onto a camera (Xenics Xeva 320). 
The complex field is retrieved using off-axis digital holography \cite{goodman1967digital, verrier2011off}.
To obtain the required holographic image (interferogram), a reference beam is separated from the main beamline with a polarizing beam splitter before the MPLC setup. 
The splitting ratio between the signal and reference beams is adjusted using a half-wave plate to obtain best interference contrast when recording the hologram.

The reference beam is magnified to a beam waist radius of $3$~mm with a beam expander 
and it's polarization is rotated by another half-wave plate to match the one of the signal beam. Then, the reference beam is overlapped with the signal beam in an off-axis configuration using a beam splitter, creating the interferogram on the camera. 
Following standard off-axis digital holography methods, the interferogram is used to retrieve the complete phase and amplitude information of the signal field by applying a 2D Fourier transform and proper filtering \cite{goodman1967digital, verrier2011off}. 
A similar interferogram is also retrieved after the MPLC and before the waveguide chip, by imaging the MPLC output on the camera with a separate 4f system 
and utilizing the same, redirected off-axis reference beam (not shown in Fig. \ref{figure1}).

\begin{figure}[htb]
\centering
\includegraphics[scale=0.55]{fig2.png}
\caption{Characterization of the mode transformation at different stages. (a) Simulated MPLC input modes $LG_{00}, LG_{10}, LG_{20}$ and the corresponding simulated waveguide modes after the transformation. (b) Experimentally reconstructed waveguide modes at the MPLC output prior to coupling to the chip. (c) Modes coupled into the waveguide chip and imaged on a camera. All field reconstructions were obtained using off-axis holography. (d) and (e) Crosstalk matrices of the transformation before and after transmission through the waveguide chip, respectively.}
\label{figure2}
\end{figure}

\section{Results}
\textbf{Coupling of ($LG_{00}$, $LG_{10}$, $LG_{20}$):}
To gauge the performance of our multi-mode interface, we test the conversion efficiency, which we define as the overlap between the experimentally realized and the simulated waveguide modes, and evaluate the visibility of these crosstalk measurements. The eigenmodes were retrieved with a finite-element simulation performed in COMSOL.
We start with a set of the three Laguerre-Gauss ($LG_{\ell,p}$) mode: $LG_{00}$, $LG_{10}$, and $LG_{20}$.
The two indices ascribed to LG modes label the azimuthal mode index $\ell$, which is connected to a helical phase $e^{i\ell\varphi}$ (azimuthal angle $\varphi$) resulting in an orbital angular momentum of $\ell$ quanta per photon, and the radial mode index $p$, which leads to radial intensity and phase profile. 

The light generated in these modes is subsequently transformed into the first three eigenmodes of the waveguide ($TE_{00}$, $TE_{10}$, $TE_{20}$) using the MPLC. 
Note, that the MPLC system works simultaneously for all modes of a given set using the same phase modulation patterns, hence its a passive interface.
To inspect the quality of the mode conversion, we retrieve the complex amplitude of the transverse light field before coupling it into the waveguide chip through recording of an interferogram and aforementioned off-axis holography.
In Fig. \ref{figure2}~a), we show the simulated free space and waveguide modes along with the experimentally retrieved waveguide modes after being converted by the MPLC in Fig. \ref{figure2}~b).  

The obtained light field resembles very well the desired higher-order modes of the waveguide chip. 
To quantify the conversion efficiency of the MPLC system, we calculate the overlap $C_{ij}$ between the simulated waveguide modes (index $i$) and the experimentally measured ones (index $j$).
The crosstalk matrix built by these overlaps is shown in Fig. \ref{figure2}~d.
The conversion efficiency between each ideal mode and its corresponding measured mode were found to be around 65$\%$ due to beam distortion, which we assume to stem from small misalignments in the MPLC system, imperfections of the phase modulations of the spatial light modulator, and imperfect imaging within the setup.
Especially the latter might have a bigger impact on our results, as they are not eigenmodes of free space propagation which results in strong beam deformation if slightly measured out of the correct imaging plane.
Nevertheless, we find that the visibility of this crosstalk matrix, defined by $V=\frac{1}{3}\Sigma_{i}\Sigma_{j} C_{ii}/C_{ij}$, remains very high at around 90\%.
The high visibility demonstrates that the imperfect transformations do not cause significant crosstalk between the generated waveguide modes.

\begin{figure}[t]
\centering
\includegraphics[scale=0.36]{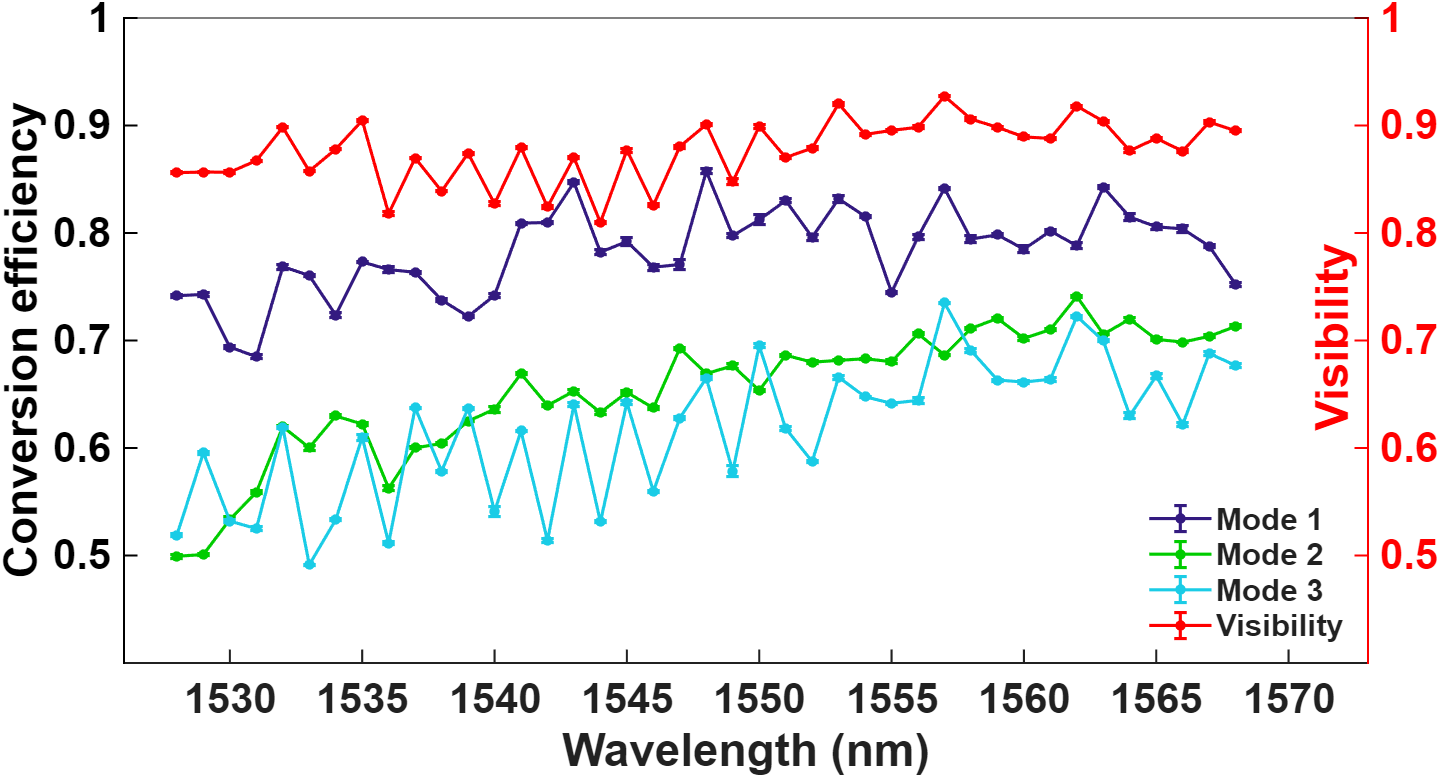}
\caption{Spectral performance of the interface. Crosstalk matrices and visibility measured after coupling to the chip for wavelengths between 1528–1568~$nm$. }
\label{figure3}
\end{figure}

The transformed light field is then coupled to the multi-mode rib waveguide. The power coupling efficiency of the generated modes to the waveguide is approximately 10--15~\%, where the value is reduced by the aforementioned issues with the MPLC and imaging systems, as well as reflections from uncoated optical surfaces.
To characterize the quality of the coupling and, thus, the overall quality of the multi-mode interface, we image the light field at the output of the waveguide chip by a camera and measure again the complex transverse amplitude using off-axis holography (see Fig. \ref{figure2}~c).
Similarly to before, we calculate the modal crosstalk matrix by comparing the measured modes with simulated ones (see Fig. \ref{figure2}~e). 
We find that the lowest order mode $TE_{00}$ has the highest conversion efficiency at 86\%, while the higher order modes decreased to around 65\%.
This difference might arise from the increased difficulty in coupling of higher-order waveguide modes, where small misalignments and imperfections have a larger effect. 
Similarly, we find a reduced overall visibility of the crosstalk matrix of around 75\%.
As there was a smaller crosstalk before the coupling to the chip and we operate with a straight waveguide that should not induce modal crosstalk, we attribute the reduction in efficiency and visibility to misalignment and imperfect imaging. 
As discussed before, especially the latter might play an important role as the waveguide modes before coupling into the chip strongly deform if slightly placed out of focus.
Nevertheless, the results indicate that our scheme is capable of efficiently interfacing light from a set of three free space LG modes ($LG_{00}$, $LG_{10}$, $LG_{20}$) to the first three higher-order waveguide modes ($TE_{00}$, $TE_{10}$, $TE_{20}$).

\textbf{Coupling for multiple wavelengths}: 
To further enhance the applicability of the interface, it will be important that it works over a broad range of different wavelengths.
For this, we included four equally-spaced wavelengths from 1540 nm to 1570 nm in the algorithm when optimizing for the four phase modulations of the MPLC, which results in smoother patterns thereby increasing the range for which the interface is working.
To experimentally verify the interface's broadband operation, we perform the above described conversion characterization, i.e., evaluating the efficiencies and visibilities from the reconstructed complex fields after the transmission through the multi-mode waveguide chips, for wavelengths between 1528 nm and 1568 nm (telecom C-band) in steps of 1 nm.
As shown in Fig. \ref{figure3}, the visibility remains fairly constant with a slight increase towards larger wavelength. 
This effect becomes even more prominent when only the overlap values between the two higher order measured and simulated modes are considered.
We attribute this effect to the larger sensitivity of shorter wavelengths with respect to misalignments and modulation imperfections of the SLM, which is optimized for an operation at 1550 nm.
Nevertheless, the results demonstrate that we can efficiently couple a set of free-space LG modes to a multi-mode waveguide across a bandwidth of approximately 40 nm (telecom C-band).\\

\begin{figure}[ht]
\centering
\includegraphics[scale=0.19]{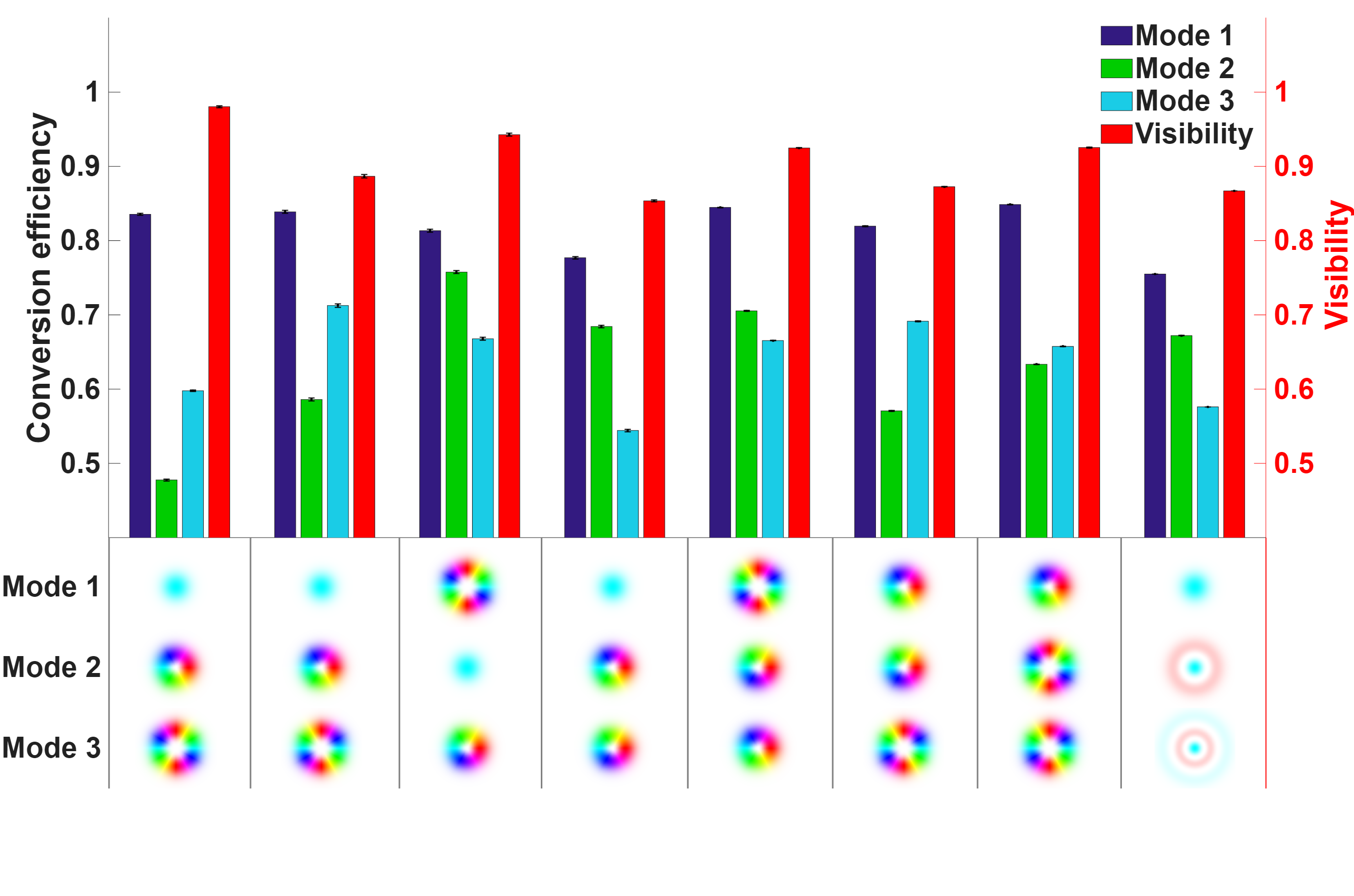}
\caption{Mode transformation with different Laguerre–Gaussian input sets.}
\label{figure4}
\end{figure}

\textbf{Coupling of various $LG$ modes:}
As a final demonstration of the capability of our multi-mode interface, we show that the interface is not limited to just the one set of modes in a coupling specific order.
We test, e.g., that the same modes as before can be coupled to same waveguide modes, however, in a different order without a loss of coupling quality.
We further study the efficiency and visibility when interfacing LG modes with different OAM values as well as modes with only radial structures.
All results are shown in Fig. \ref{figure4}. 
Note that for each set of modes, we optimized the phase modulations of the MPLC.
As before, we then imprinted these modes on a light field, transformed the light via the MPLC, coupled it into the waveguide, and recorded the emitted light field after the waveguide chip to calculate the modal efficiencies and crosstalk.
We consistently achieved efficiency values between 60\% and 85\% with visibilities that vary between 70\% and 90\% for all transformations demonstrating that our scheme is capable to interface a large variety of higher order modes.
Hence, we anticipate that it will not only be able to interface free-space light fields with photonic integrated circuits but also similar higher-order modes of other systems such as multi-mode fibers.

\section{Conclusion}
Connecting different multi-mode encoding schemes and platforms of optical communications will be essential in future high-capacity, high-speed optical networks and optical information processing. 
Our work demonstrates an efficient, versatile system to connect free-space with integrated multi-mode optical systems.
We show that a large variety of different sets of Laguerre-Gauss modes can be transformed to and coupled into multimode waveguides of a photonic integrated circuit. 
As such its shows a promising way forward to increase data capacities in terms of multiplexing information via higher-order light modes.
Being a passive modulation scheme that works over a broad range of wavelengths (here the entire telecom C-band) the interface will not impose any limitations on already achievable data rates even if frequency-multiplexing is applied.
In addition, slight modifications to the MPLC scheme, e.g., through the use of polarization-modulating metasurfaces \cite{soma2025complete}, will allow to extend the scheme to also include polarization-multiplexed signal.
Finally, we anticipate that in the future more compact and efficient schemes such a larger number of modes with increased conversion effiency and reduced crosstalk will be possible \cite{fang2021performance,MPLc1,MPLc2}.

\section*{Acknowledgment} 
TS, OK, ME, RF acknowledge the support of the Research Council of Finland (RCF) through the Photonics Research and Innovation Flagship PREIN (decision 320165). TS and OK acknowledge the RCF through the project BIQOS (decision 358134). ME and RF acknowledge the European Research Council project TWISTION (decision 101042368). RF acknowledges TCF through the Academy Research Fellowship (decision 332399). MC and GT, thank funding from PREIN (Decision No. 320168). GT also acknowledges support from RCF through the project CryOpto (Decision No.  363464).

\section*{Disclosures} 
The authors declare no conflicts of interest.
\section*{Data availability}
Data underlying the results presented in this paper may be obtained from the authors upon reasonable request.
\bibliography{sample}

\end{document}